\documentclass[onecolumn,10p,secnumarabic,a4paper,nobibnotes,aps,prd,showpacs,showkeys,notitlepage]{revtex4-1}
\usepackage[T1]{fontenc}
\usepackage{amsmath}
\usepackage{amssymb}
\usepackage[dvipdfm]{epsfig}
\usepackage{graphicx}
\usepackage{natbib}
\usepackage[margin=4cm]{geometry}

\begin{document}

\title{An Approach to Modeling and Scaling of Hysteresis in Soft Magnetic Materials. I Magnetization Curve }  

\author{Krzysztof Z. Sokalski}

\affiliation{Institute of Computer Science, 
Cz\c{e}stochowa University of Technology, 
Al. Armii Krajowej 17, 42-200 Cz\c{e}stochowa, Poland}

\keywords{ Mathematical model of magnetization, hysteresis loops,scaling}

\pacs{75.50.-y, 75.60.-d, 89.75.Da}


\begin{abstract}
A new mathematical model of hysteresis loop has been derived. Model consists in an extansion of tanh($\cdot$) by extanding the base of exp function into an arbitrary positive number. The presented model is self-similar and invariant with respect to scaling.  Scaling of magnetic hysteresis loop has been done using the notion of homogenous function in general sense. 
\end{abstract}

\maketitle

\section{Introduction}\label{I}
  In the two last decade the following models have been accepted for a proper description of the hysteresis loop: 
1) Jiles-Atherton \citep{bib:JA1}, \citep{bib:JA2} and Szczyg{\l}owski's supplement \citep{bib:SZCZ} as well as Chwastek's modification \citep{bib:CHW},  
2) Harrison \citep{bib:HAR1}, \citep{bib:HAR2}, 
3) Tak\'{a}cs \citep{bib:TAK1} , \citep{bib:TAK2}, 
4) Zirka-Moroz \citep{bib:ZIMO}.\\ 
 For an older approaches to the problem of the magnetization hysteresis we refer to books by Bertotti \citep{bib:GB} and Cullity \citep{bib:Cull}. For more recent approaches we refer to the monograph \citep{bib:BEMA}.
Despite the multiplicity of hysteresis models developed to date, there is no model of which self-similarity can be expressed by homogeneous functions in general sense. While the scaling of power losses is already solved by the notion of function heaving that property \citep{bib:SOK1}. Since these two problems are related by the surface of hysteresis loop, it should exist an analogical representation of scaling for the magnetization processes. Goal of the presented paper is to create a mathematical model which describes hysteresis and enables to express its self-similarity by the homogeneous function in general sense. This function should enable us to reproduce all magnetization processes inside a major loop which will play a role of an envelop for these processes.\\
It is well known that tanh($\cdot$) suits for model of initial magnetization function. It describes properly the saturations for both asymptotic values of the magnetic field: $H\rightarrow \pm\infty$ as well  as behavior of the magnetization in the neighborhood of origin. The mentioned above paper \citep{bib:TAK1} starts from tanh($\cdot$) and describes phenomenological model which gives complete description of the magnetic hysteresis. Moreover, this mapping helps to solve problems of modeling of core losses in the present of DC bias by the projection of the magnetic field just into the initial curve \citep{bib:Rusz1}.  
In this paper we present Mathematical Model of hysteresis (MMH) in Soft Magnetic Materials which bases on an extension of the function tanh($\cdot$). Goal of this paper is to derive  mathematical model which properly describes the magnetization hysteresis and it enables us to perform scaling of the hysteresis loop basing on  notions of  self-similar system and  homogeneous function in general sense \citep{bib:BAR}, \citep{bib:SOK1}. This  paper is organized as following. In section \ref{I} we introduce the extend tanh($\cdot$) function from which 
we derive its basic properties. In section \ref{II} we deal with self-similarity of the considered model and scaling. Conclusions are presented in Section \ref{CON}.
\section{Extension of tanh$({\cdot})$ function and  Mathematical Model of hysteresis}\label{II}

Let us start from the definition of tanh($\cdot$):
\begin{equation}
\label{tynh}
tanh(x)=\frac{e^{x}-e^{-x}}{e^{x}+e^{-x}},
\end{equation}
where $e$ is the base of the so-called natural logarithm. So, let us generalize (\ref{tynh}) by introducing the four bases instead $e$:
 \begin{equation}
\label{tynh1}
tanH(a,b,c,d|x)=\frac{a^{x}-b^{-x}}{c^{x}+d^{-x}},
\end{equation}
where $a,b,c,d$ are arbitrary positive numbers. 
\subsection{Hysteresis in  $[$Magnetic Field, Magnetization$]$ plane}
In the first step we will derive a model of major hysteresis loop which is an envelope ot the whole family of loops of the particular considered case. 
Let us write down the model expression for initial magnetization curve:
\begin{equation}
\label{tynh2}
M_{P}(X)=M_{0}\,P(X,\epsilon); \hspace{2mm}X\in[0,X_{max}],
\end{equation}
where, $M_{0}$ is magnetization corresponding to saturation expressed in Tesla: $[$T$]$,  $X\in \{0,X_{max}\}$, $X=\frac{H}{h}$ where $H$ is magnetic field, $h$ is a parameter of the magnetic field dimension $[$A m$^{-1}]$ to be determined. where, function $P(X,\epsilon)$ is of the following form: 
\begin{equation}
\label{primary}
P(X,\epsilon)=\frac {{a}^{X-\epsilon}-{b}^{-X+\epsilon}}{{c}^{X-\epsilon}+{d}^{-X+\epsilon}},
\end{equation}
where $\epsilon$ is  modeling parameter of the order $\theta/2$.
Let the upper section of hysteresis loop and the lower one are of the following forms:
\begin{equation}
\label{major}
M_{F}(X)=M_{0}F(X,\theta); \hspace{2mm}M_{G}(X)=M_{0}G(X,\theta),
\end{equation}
where
\begin{eqnarray}
F(X,\theta)=\frac {{a}^{X+\theta}-{b}^{-X-\theta}}{{c}^{X+\theta}+{d}^{-X-\theta}}\label{d},\\
G(X,\theta)=\frac {{a}^{X-\theta}-{b}^{-X+\theta}}{{c}^{X-\theta}+{d}^{-X+\theta}}\label{g}.
\end{eqnarray}
 $\theta$ is a model parameter depending on $X_{max}$. 

.
Let us consider for illustration the following example: $M_{0} = 1; a = 4; b = 4; c = 4; d = 4; \theta_{i} = 1.3$ for $i=a,b,c,d$.
\begin{figure}
\begin{center}
\includegraphics[ width=10cm]{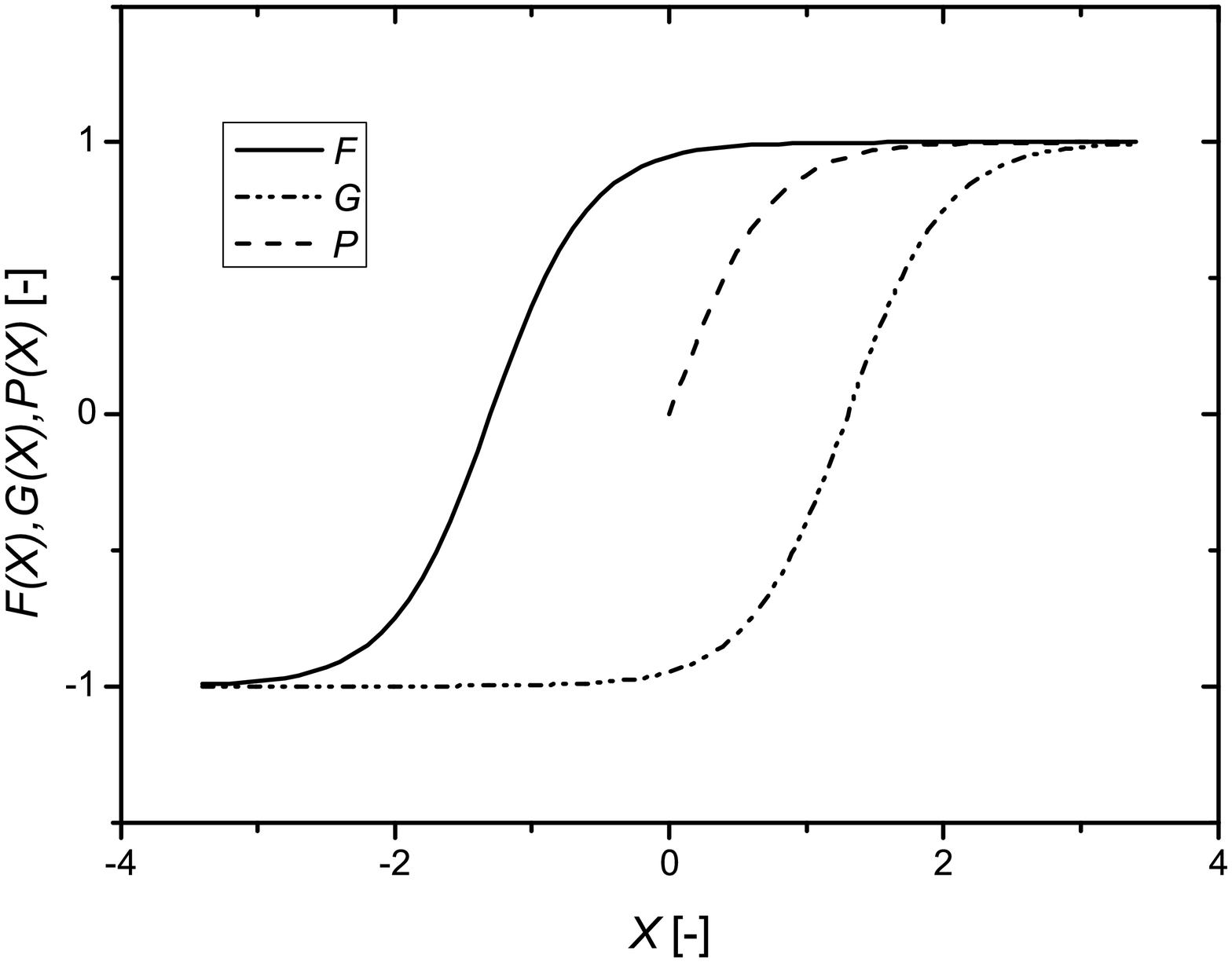}
\caption{A model of nucleation type hysteresis constructed with functions $F$, $P$ and $G$ according  to (\ref{d}), (\ref{g}) and (\ref{primary})}
\label{Fig.3}
\end{center}
\end{figure} 
The magnetization as a function of the magnetic field possesses  horizontal asymptotes for $H\rightarrow \pm \infty$. This means for our modeling, that the functions  $F(X,\theta)$ and $G(X,\theta)$ have to possess the same asymptotic properties. These components get  equal values for $|H|=\infty$. In fact due to uncertainty of meassured magnitudes it is possible to accept the saturation points at $X=X_{min}$ and $X=X_{max}$ being the end points of the hysteresis loop. Therefore,  the modeling process has to ensure that the initial function $P(H)$ matches the origin and the upper end point of the loop. This effect can be controlled by the $\theta$ parameter Figure \ref{Fig.3}. Values of $X_{max}$ is determined by the following relation:
\begin {equation}
\label{uncertain}
|F(X_{max},\theta)-G(X_{max},\theta)|\leq |\psi|,
\end{equation}
where $\psi$ is measure of uncertainty of $|M_{F}(X)-M_{G}(X)|$ in dimension less units: $\psi=\delta |M_{F}(X)-M_{G}(X)|/M_{0}$.

\subsection{Scaling of magnetization's hysteresis loop }
We present the scaling procedure on  example of the both sections of a major loop. Functions which have to be scaled (\ref{primary}),(\ref{d}),(\ref{g}) consist of exponential functions and their arguments are exponents. In such a case the scaling can be performed on the basis of these functions, while the exponents can be converted by gauge transformations. 
 Let us assume that there exist real numbers $\{\alpha,\beta,\gamma,\delta, \nu \}\in \mathbf{R}^{5}$ such that $\forall{\lambda}\in \mathbf{R}_{+}$ and $\forall{\chi}\in \mathbf{R}$ the following relations hold:
\begin{eqnarray}
\frac{M_{F}(X)}{M_{0}}\lambda^{\nu}=\frac{(\lambda^{\alpha}a)^{X+\theta}a^{\chi}-(\lambda^{\beta}b)^{-X-\theta}a^{-\chi}}{(\lambda^{\gamma}c)^{X+\theta}c^{\chi}+(\lambda^{\delta}d)^{-X-\theta}d^{-\chi}},\nonumber\\
\frac{M_{G}(X)}{M_{0}}\lambda^{\nu}=\frac{(\lambda^{\alpha}a)^{X-\theta}a^{\chi}-(\lambda^{\beta}b)^{-X+\theta}b^{-\chi}}{(\lambda^{\gamma}c)^{X-\theta}c^{\chi}+(\lambda^{\delta}d)^{-X+\theta}d^{-\chi}},\label{scaling1}
\end{eqnarray}
then $M_{F}(X)$ and $M_{G}(X)$ is set of homogenous functions in general sense.
\subsection{Group symmetry of the considered model}
 $\alpha,\beta,\gamma,\delta, \nu$ are scaling exponents and $\chi$ is a gauge transformation parameter. We have assumed that the gauge transformation acted directly on the argument of the magnetization function: $M_{F}(X)\rightarrow M_{F}(X+\chi)$. Whereas scaling acts directly on each base of exponential functions, for example: $a^{X}\rightarrow (\lambda^{\alpha}a)^{X}$. We will show that  this action can be transformed to an operator acting directly on the linear function of $X$.
The derived transformation (\ref{scaling1}) depends on the two parameters: $\lambda$ and $\chi$. Each of them generates one parameter group: multiplicative group   $\mathcal{G}_{\lambda}$ and additive one $\mathcal{G}_{\chi}$, repectively. By the definitions how do the scaling and gauge transformations act on the model equations the group of symmetry of (\ref{scaling1}) is the following semidirect product \citep{bib:Hamm}:
\begin{equation}
\label{group1}
\mathcal{G}_{\lambda,\chi}=\mathcal{G}_{\lambda} \rtimes \mathcal{G}_{\chi}.
\end{equation}.
Group action in $\mathcal{G}_{\lambda,\chi}$ is defined as following. Let $\{p_{1},\chi_{1}\}\in \mathcal{G}_{\lambda,\chi}$ and $\{p_{2},\chi_{2}\}\in \mathcal{G}_{\lambda,\chi}$. Then:
\begin{equation}
\label{action}
\{p_{2},\chi_{2}\}\rtimes \{p_{1},\chi_{1}\}=\{p_{2}p_{1},p_{2}\chi_{1}+\chi_{2}\}.
\end{equation}
The group structure of $\mathcal{G}_{\lambda,\chi}$ will play important role in derivation relations between hysteresis loops.
\subsection{Symmetric model} 
Now we are ready to prove that the symmetric model of the hysteresis loop is invariant with respect to ${\lambda_{i},\chi_{j}}\in \mathcal{G}_{\lambda,\chi}$. By the symmetric model we mean a case characterized by the following relations between basis of (\ref{primary}),(\ref{d}),(\ref{g}):
\begin{equation}
\label{abcd}
a=b=c=d.
\end{equation}.
Otherwise a model is non-symmetric. For further considerations we take into account the following formula for $M_{F}(X),M_{G}(X),M_{P}(X)$:
\begin{eqnarray}
\frac{M_{F}(X)}{M_{0}}=\frac{(a)^{X+\theta}-(a)^{-X-\theta}}{( a)^{X+\theta}+(a)^{-X-\theta}},\label{12a}\\
\frac{M_{G}(X)}{M_{0}}=\frac{(a)^{X-\theta}-(a)^{-X+\theta}}{( a)^{X-\theta}+(a)^{-X+\theta}},\label{12b}\\
\frac{M_{P}(X)}{M_{0}}=\frac{(a)^{X+\epsilon}-(a)^{-X-\epsilon}}{( a)^{X+\epsilon}+(a)^{-X-\epsilon}}\label{12c}
\end{eqnarray}

{\em Theorem}: If basis of the model (\ref{primary}), (\ref{d}) and (\ref{g}) satisfy (\ref{abcd}) then the model of hysteresis loop is invariant with respect to scaling and gauge transformation.\\
{\em Proof}: Let us assume an action of a group element belonging to $\mathcal{G}_{\lambda,\chi}$. For the symmetry condition (\ref{abcd}) the relation (\ref{scaling1}) takes the following form:
\begin{eqnarray}
\frac{M_{F}(X)}{M_{0}}\lambda^{\nu}=\frac{(\lambda^{\alpha}a)^{X+\theta}a^{\chi}-(\lambda^{\alpha}a)^{-X-\theta}a^{-\chi}}{(\lambda^{\alpha}a)^{X+\theta}a^{\chi}+(\lambda^{\alpha}a)^{-X-\theta}a^{-\chi}},\nonumber\\
\frac{M_{G}(X)}{M_{0}}\lambda^{\nu}=\frac{(\lambda^{\alpha}a)^{X-\theta}a^{\chi}-(\lambda^{\alpha}a)^{-X+\theta}a^{-\chi}}{(\lambda^{\alpha}a)^{X-\theta}a^{\chi}+(\lambda^{\alpha}a)^{-X+\theta}a^{-\chi}},\label{scaling2}
\end{eqnarray}
where we have to remember that according to (\ref{action}) the order of action is relevant. First the scaling has to be done before the gauge transformation.
According to the assumtions just above (\ref{scaling1}) we are free to asume that 
\begin{equation}
\label{scaling2p}
\lambda^{\alpha}=a^{p-1}. 
\end{equation}
Then the relations (\ref{scaling2}) after simple evaluations take the following form:
\begin{eqnarray}
\frac{M_{F}(X)}{M_{0}}a^{n}=\frac{(a)^{p\,(X+\theta)+\chi}-(a)^{p\,(-X-\theta)-\chi}}{(a)^{p\,(X+\theta)+\chi}+(a)^{p\,(-X-\theta)-\chi}},\nonumber\\
\frac{M_{G}(X)}{M_{0}}a^{n}=\frac{(a)^{p\,(X-\theta)+\chi}-(a)^{p\,(-X+\theta)-\chi}}{(a)^{p\,(X-\theta)+\chi}+(a)^{p\,(-X+\theta)-\chi}},\label{scaling3}
\end{eqnarray}
where $n=\frac{\nu}{\alpha}(p-1)$ and $p\in \{\mathbf{R}/0\}$.
Introducing the following new variables: 
\begin{equation}
\label{scaling4}
M_{0}'=M_{0}a^{-n},\hspace{2mm} X'=p\,X+\chi,\hspace{2mm}\theta'=p\,\theta
\end{equation}
 we get (\ref{12a}).\\
The initial magnetization curve (\ref{12c}) is invariant with respect to the scaling and the gauge transformations, also. The proof goes by the same way as for (\ref{12a}) and (\ref{12b}).
 $\Box$ 
\subsection{Interpretation of scaling and gauge transformation}
\begin{figure}
\begin{center}
\includegraphics[ width=10cm]{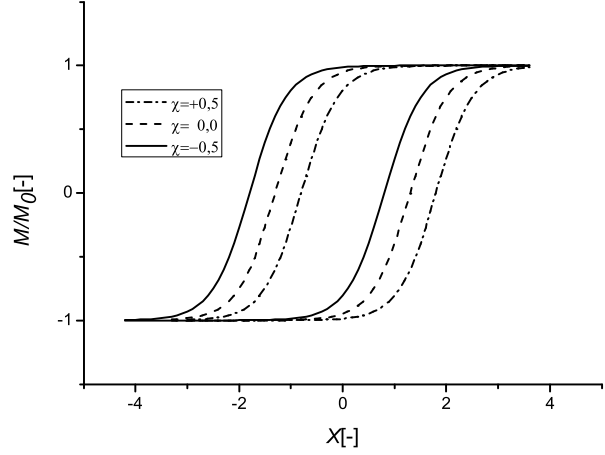}
\caption{Magnetic hysteresis family for $p=1\hspace{2mm}n=1\hspace{2mm}\theta=1,3\hspace{2mm}\nu/\alpha=1$.}
\label{Fig.4}
\end{center}
\end{figure}

\begin{figure}
\begin{center}
\includegraphics[ width=10cm]{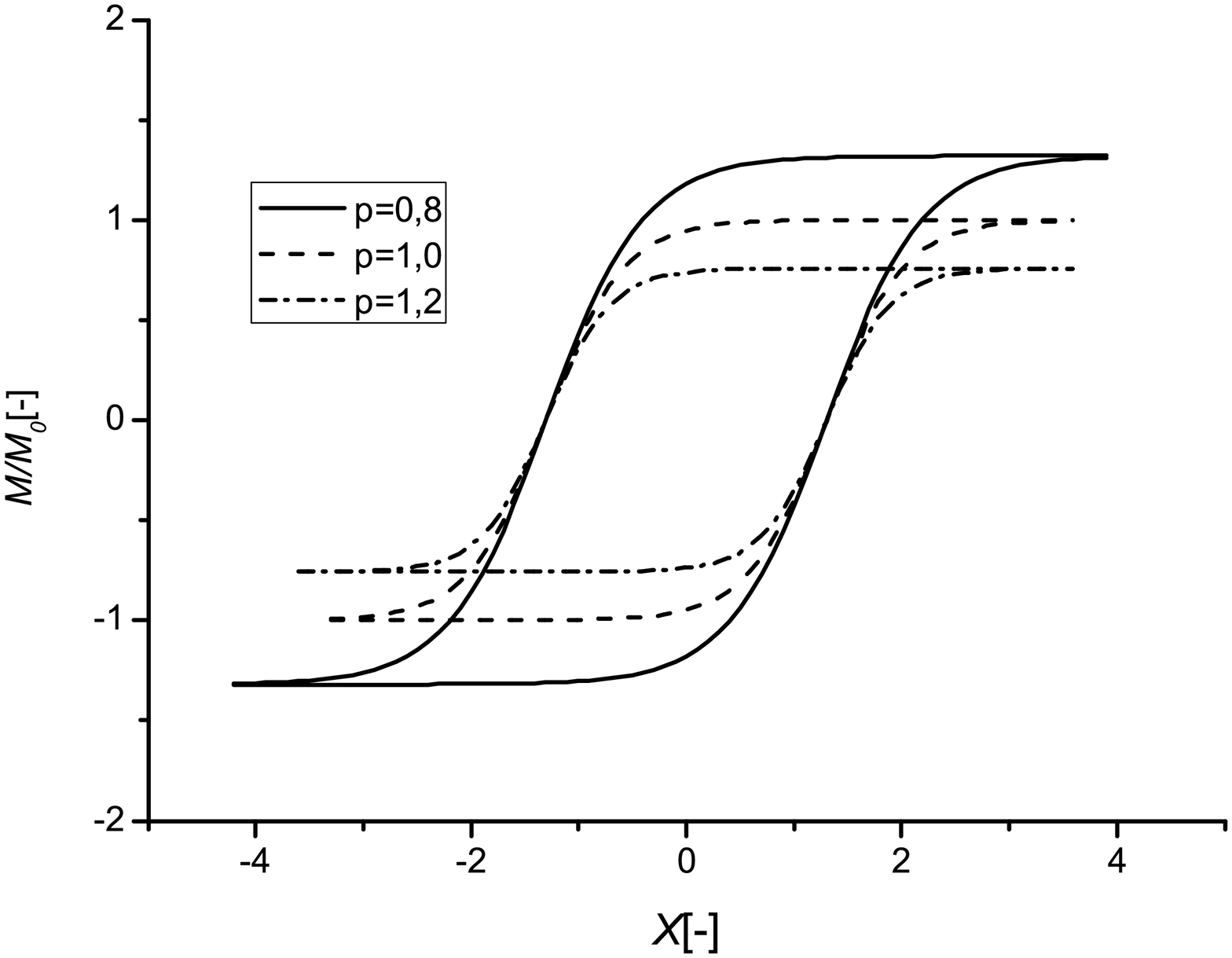}
\caption{Magnetic hysteresis family for $\chi=0\hspace{2mm}n=1\hspace{2mm}\theta=1,3\hspace{2mm}\nu/\alpha=1.$}
\label{Fig.5}
\end{center}
\end{figure}

\begin{figure}
\begin{center}
\includegraphics[ width=10cm]{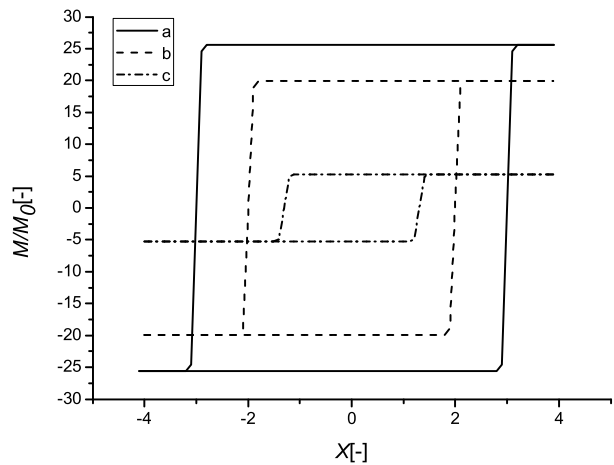}
\caption{Magnetic hysteresis family for the following values of the model and scaling parameters: a) $\theta=3\hspace{2mm}p=14\hspace{2mm}\nu/\alpha=-0,18\hspace{2mm}\chi=0$,
b) $\theta=2\hspace{2mm}p=13\hspace{2mm}\nu/\alpha=-0,18\hspace{2mm}\chi=0$, c) $\theta=1,3\hspace{2mm}p=13\hspace{2mm}\nu/\alpha=0\hspace{2mm}\chi=0$.}
\label{Fig.6}
\end{center}
\end{figure}

In order to make an interpretation of the derived scaling and gauge transformation we present results of their action on the hysteresis models Figure \ref{Fig.4}-Figure \ref{Fig.6}.  Figure \ref{Fig.4} presents how  pure gauge transformations generate a displacement of transformed loops along the horizontal axis. Figure \ref{Fig.5} presents compression of a loop along the vertical axis under the scaling. Finally, for large value of the scaling parameter $p$ the transformed loops become to the Preisach classical models (Figure \ref{Fig.6} \citep{bib:Pre}).   
 \section{Conclusions}\label{CON}
By physical models of the hysteresis loops we try to explain origin of the magnetic materials properties. Whereas in designing the  magnetic materials we need  analytic formulas returning real values of physical magnitudes which are relevant in this process. Just the mathematical models together with an experimental data constitute powerful tool for designers \citep{bib:SLUS},\citep{bib:SOKNEW}. 
In this paper we have derived the simple model which is promising for the mentioned above applications. The introduced group theoretical methods in analysis of the hysteresis loops will help us to divide the hysteresis loops space into orbits in $\mathcal{G}_{\lambda,\chi}$.  This division will enable us to create a minimal set of independent loops which will constitute a base in the hysteresis loops space.

\bibliographystyle{plainnat}
 
\end{document}